\documentclass[runningheads]{llncs}
\usepackage[T1]{fontenc}
\usepackage{graphicx}
\usepackage{amsmath}
\usepackage{orcidlink}
\usepackage{hyperref}
\usepackage{enumitem}
\usepackage{amsfonts}
\usepackage{multirow} 
\usepackage{subcaption}
\usepackage{xcolor}
\usepackage{float} 
\usepackage{booktabs}%
\usepackage{listings}
\usepackage{csquotes}

\usepackage{amsthm}

\usepackage{tikz}
\usepackage{tcolorbox}
\newtheoremstyle{styledef}%
{3pt}
{3pt}
{}
{}
{\bfseries\color{black}}
{}
{.5em}
{}
\theoremstyle{styledef}

\tcolorboxenvironment{definition}{
  colback=lightgray!25, 
  colframe=lightgray!25, 
  fonttitle=\bfseries, 
  boxrule=0.5pt, 
  coltitle=black, 
  rounded corners, 
  left=5pt, 
  right=5pt, 
  top=5pt, 
  bottom=5pt, 
}

\tcolorboxenvironment{example}{
  colback=lightgray!25, 
  colframe=lightgray!25, 
  fonttitle=\bfseries, 
  boxrule=0.5pt, 
  coltitle=black, 
  rounded corners, 
  left=5pt, 
  right=5pt, 
  top=5pt, 
  bottom=5pt, 
}

\definecolor{bg}{gray}{0.92} 
\definecolor{keywordcolor}{HTML}{569CD6}
\definecolor{stringcolor}{HTML}{8E4A3D}
\definecolor{typecolor}{rgb}{0.0, 0.42, 0.24}
\definecolor{objectcolor}{RGB}{0, 51, 102} 
\definecolor{functioncolor}{rgb}{0.72, 0.53, 0.04} 
\definecolor{commentcolor}{HTML}{6A9955}
\definecolor{prefixcolor}{HTML}{6A9955}
\definecolor{defaultcolor}{HTML}{000000}
\definecolor{framegray}{gray}{0.6}
\definecolor{datastructurecolor}{HTML}{FF6F61}

\definecolor{mygreen}{rgb}{0,0.5,0}      
\definecolor{myblue}{rgb}{0,0,0.75}      
\definecolor{mypurple}{rgb}{0.5,0,0.5}   
\definecolor{mygray}{rgb}{0.5,0.5,0.5}   
\lstdefinestyle{vscodestyle}{
    backgroundcolor=\color{bg},
    basicstyle=\ttfamily\small\color{defaultcolor},
    keywordstyle=\color{keywordcolor}\bfseries,
    stringstyle=\color{stringcolor}\bfseries,
    commentstyle=\color{commentcolor}\itshape,
    morecomment=[l]{//},
    morecomment=[s]{/*}{*/},
    morekeywords={const, let, var, new, function, return, if, else, JSMF, for, print, fs, on, pipe, reject, row, parseFloat, forEach, console, true, FILTER, BIND, CONCAT, STR, CONTAINS, AS},
    showstringspaces=false,
    breaklines=true,
    columns=fullflexible,
    keepspaces=true,
    frame=single,
    rulecolor=\color{framegray},
    xleftmargin=1em,
    morestring=[b]',         
    morestring=[b]",         
    emph={String, Number, Boolean, Object, Array}, 
    emphstyle=\color{typecolor},
    emph={[2]newInstance,addAttribute,setReference, setDescription, setModellingElements, items, resolve, setSemanticReferences, log, setSemanticReference},
    emphstyle={[2]\color{functioncolor}},
    emph={[3] Building, Room, Controller, Alarm, HumSensor, TempSensor, TemperatureSensor, HumiditySensor, Proximity, PressureSensor, ProxSensor, airquality,  Model, Class, building, room, controller,controller1, tempSensor, humSensor, proxSensor, alarm, airquality_model, instances, modelElements, element, semanticMappings,top_matches, setFlexible, MM, M, dt, mm, brick, owl, ts, hs, sensorType, rdf},
    emphstyle={[3]\color{objectcolor}\bfseries},
    emph={[4] PREFIX, SELECT, WHERE},
    emphstyle={[4]\color{mygreen}\bfseries}
}

\begin{document}
\title{Semantic Grounding of Digital Twin Metamodels Using RDF Graphs}
\author{Faima Abbasi\inst{1,2}\orcidlink{0009-0001-7484-9256} \and
Jean-Sébastien Sottet\inst{1}\orcidlink{0000-0002-3071-6371} \and
Cedric Pruski\inst{1}\orcidlink{0000-0002-2103-0431}}
\authorrunning{F. Abbasi et al.}
\institute{Luxembourg Institute of Science and Technology, 5 Avenue des Hauts-Fourneaux, L-4362 Esch-sur-Alzette, Luxembourg\and
FSTM, University of Luxembourg, 2 Av. de l'Universite, L-4365 Esch-sur-Alzette, Luxembourg\\
\email{\{faima.abbasi, jean-sebastien.sottet, cedric.pruski\}@list.lu}}
\maketitle 
\begin{abstract}
Digital Twins (DTs) represent digital counterparts of physical systems, assets, or processes, referred to as the actual twin (AT). DTs integrate heterogeneous data, models, and semantic technologies to support monitoring, simulation, prediction, and optimization, enabling informed decision-making while maintaining a dynamic and accurate reflection of the AT. A key challenge is aligning heterogeneous models, which can cause semantic mismatches, inconsistencies, and synchronization issues. Existing approaches relying on static mappings and manual updates are often inflexible and error-prone. In this study, we address heterogeneity challenge in multi-layered DT, by introducing semantic grounding pipeline for multi-layered DTs that enables consistent and reliable interoperability between abstraction layers. We make three contributions. First, we design and implement multi-layered DT using flexible modelling framework, to organize data, model and metamodel layers. Second, we semantically lift DT metamodel to \texttt{RDF} graph for unified representation. Finally, we present a graph-based alignment approach (\texttt{SSM-OM}), which leverages semantic embeddings, lexical similarity, and large language model (\texttt{LLM}) reasoning to accurately establish and validate correspondences between the lifted metamodel and ontology. We validate correctness, interoperability, cross-layer traceability, domain applicability and general empirical performance through \texttt{RDF} tests, a DT usecase, and ontology alignment evaluation initiative (\texttt{OAEI}\footnote{\url{https://oaei.ontologymatching.org/}})
benchmarks, demonstrating semantic consistency in multi-layered DT.

\end{abstract}

\section{Introduction} \label{sec1}
DTs connect AT with their digital models through a continuous and closed feedback loop, enabling clear, real-time monitoring and analysis throughout the lifecycle of a system \cite{gil2025architecture}. DTs are complex software systems composed of diverse artefacts that vary by usecase, requiring significant development effort and close collaboration between software and domain experts \cite{lehner2025model,lehner2023architectural,dalibor2022cross}.
\begin{figure}[h]
\centering
\includegraphics[width=3.9in, height=2in]{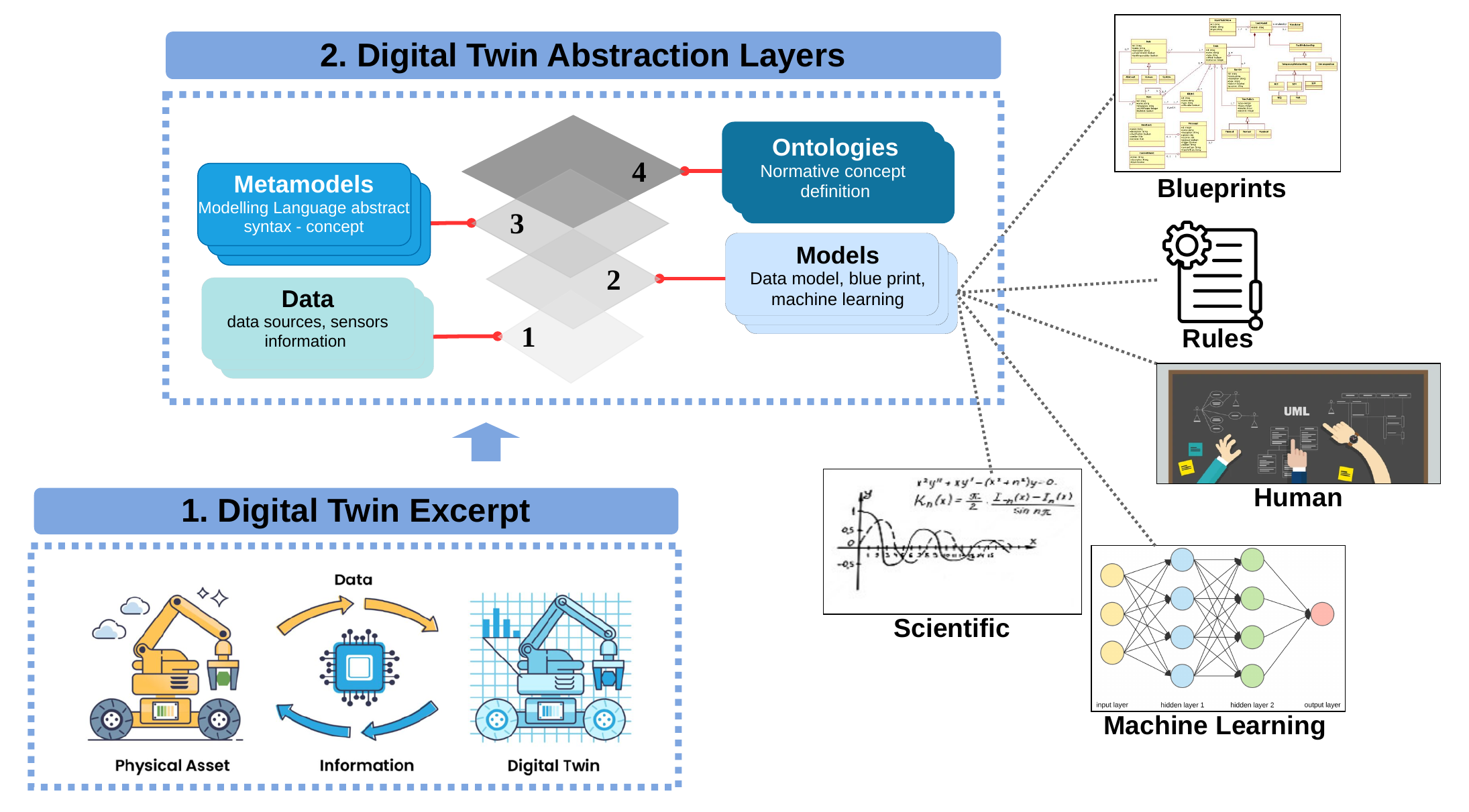}
\caption{An Illustration of DT Abstraction Layers}
\label{Fig1}
\end{figure}
Model-driven engineering (MDE) manages complexity through abstract models that are human-readable and machine-executable \cite{brambilla2017model}. DTs depend on multiple heterogeneous models for descriptive, prescriptive and predictive purposes \cite{bordeleau2020towards}, which must be coordinated across engineering domains to support their development, operation, and evolution \cite{dalibor2022cross,lehner2025model,sottet2025leverage}. DT lifecycle spans on design and creation, manufacturing or production, operation, and end-of-life phases, reflecting its roots in product lifecycle management and enabling continuous integration of data across all stages\cite{pronost2024digital}. However, it requires a semantic extension to ensure interoperability and consistency. This should include semantic grounding concepts in pipeline, i.e., \textbf{aligning} DT metamodels with domain ontologies during design, \textbf{mapping} sensor and operational data to semantic concepts during deployment, \textbf{evolving} and maintaining these mappings throughout operation, and enabling \textbf{reasoning} and querying across abstraction layers during operation. Such an extension can be seen as an additional layer complementing existing modelling and data-driven approaches in DTs, where ontologies and knowledge graphs support semantic interoperability and informed decision-making \cite{karabulut2024ontologies,sottet2025leverage}. 
In our previous work \cite{abbasi2024understanding,abbasi2025semantic}, we identified four levels of abstraction in multi-layered DT (see Fig. \ref{Fig1}), which can be mapped to system design and creation phase of DT life cycle, i.e., (i) \textbf{data:} represents information related to data sources, (ii) \textbf{models:} represents an abstract system, (iii) \textbf{metamodels/schema:} uses a specific domain to define the structure and semantics of models, and (iv) \textbf{ontologies:} at a higher level, they represent the shared understanding of concepts and relationships within a domain. These elements constitute heterogeneous models, as they differ in abstraction level, representation, and purpose, ranging from structural definitions to semantic descriptions. While these abstraction levels evolve independently, aligning and integrating these heterogeneous artifacts is therefore crucial to maintain semantic coherence, interoperability, cross-layer reasoning and reliability of DT across its life cycle. A formal alignment function is required to map metamodel entities to semantically equivalent ontology concepts, enabling semantic grounding. Semantics propagate downward: model elements derive their meaning through conformance to the metamodel, while data instances inherit semantics through their instantiation of model elements. This layered propagation ensures integration of heterogeneous artifacts, semantic coherence, interoperability, and reliable DT behavior throughout its lifecycle. Recent works have advanced DT model alignment \cite{su2025digital,zheng2020generic,franco2025multi}, yet a unified mechanism across all four layers, i.e., data, models, metamodels/schemas, and ontologies, remains lacking. Heterogeneous artifacts rely on different formalisms (e.g., \texttt{JSON}-based metamodels, \texttt{UML}, \texttt{OWL}), complicating direct alignment.  
Metamodels can be expressed in \texttt{RDF} as a common standard, serving as a pivot compatible with ontologies and knowledge graphs. During its lifecycle DT must align, map, maintain, reason and evolve its heterogeneous models to ensure semantic coherence, interoperability, and reliability. The challenges remain in preserving consistency, managing heterogeneity, and defining general validation methods.

In this work, we tackle the heterogeneity of modeling artifacts in multi-layered DTs by introducing a semantic grounding pipeline, with a particular focus on its alignment and mapping steps. It is built on three core concepts: (i) multi-layered DT design and implementation using a flexible modeling framework to systematically organize data, models, and metamodels; (ii) semantic lifting, transforming DT metamodels into \texttt{RDF} graphs for unified representation and alignment; and (iii) DT-specific alignment, combining embeddings, lexical similarity and \texttt{LLM} reasoning to establish, refine and validate correspondences against domain knowledge. Although the pipeline covers all phases of semantic grounding, including reasoning and evolution, this study is focused on alignment and mapping. The approach ensures semantic grounding and structural consistency, supporting robust multi-layer DT engineering rather than standalone ontology matching.

Our key contributions are: (i) we emphasis lifting heterogeneous metamodels into \texttt{RDF} graphs to unify their representation and enable semantic grounding, (ii) we perform graph-based alignment between lifted metamodel and ontology, proposing a semantics and structure-aware metamodel ontology matching (\texttt{SSM-OM}) method, and (iii) we conduct an empirical evaluation using a DT use case to assess interoperability, cross-layer traceability and domain applicability, complemented by benchmark test cases from \texttt{OAEI} to evaluate general alignment performance, along with validation of the correctness of the \texttt{RDF} transformation. Reproducibility of experiments can be found here \footnote{\url{https://github.com/faimaAbbasi/dt-metamodel-semantic-grounding.git}}. 

Rest of the paper covers:  related work in Section \ref{sec2}, DT semantic grounding in Section \ref{sec3}, indoor air quality usecase in Section \ref{sec4}, graph-based alignment approach in Section \ref{sec5}, evaluation in Section \ref{sec6}, followed by discussion and conclusion in Section \ref{sec7} and \ref{sec8}.

\section{Related Work}\label{sec2}
We categorize related work into multi-layer DT alignment, semantic lifting, ontology and knowledge graph alignment, highlighting key approaches, gaps, and motivation. 

\textit{Multi-layered DT Alignment} We review studies related to multi-layered DT alignment, noting that most of the research focuses on model–model alignment, with limited attention to model–metamodel and metamodel–ontology alignment. Model alignment ensures consistency across or within abstraction layers by linking equivalent elements. Su et al. \cite{su2025digital} propose a three-layer knowledge graph architecture: (i) a concept layer organizing key information, (ii) a model layer aligning digital and physical parameters, and (iii) a decision layer integrating models with real-time data. This structure supports model alignment through knowledge graph and ensures cross-layer consistency, validated in aero-engine blade production through multi-source data integration, improving prediction, anomaly detection, and process control. Woitsch et al. \cite{woitsch2022model} present a model-driven approach for lifecycle wide data integration, aligning product, process, and service models through consistent interfaces and transformation rules to ensure semantic coherence and interoperability. This enables real-time data exchange and synchronization between AT and DT models, supported by ontologies and model transformations for dynamic updates and lifecycle optimization. Franco et al. \cite{franco2025multi} propose a five-layer human DT architecture: (i) physical, (ii) data, (iii) model, (iv) integration, and (v) interaction. Their modular design enables model–model alignment through standardized interfaces and protocols, ensuring consistent data and behavior synchronization across system. Zheng et al. \cite{zheng2020generic} introduce a tri-model CPS architecture (digital, computational, graph-based) that supports model–model matching and seamless data flow through a digital thread, ensuring real-time consistency and enhancing DT interoperability and adaptability in manufacturing systems.

\textit{Semantic Lifting} This concept is explored in various forms in the literature \cite{kappel2006lifting,kamburjan2024semantic,krauter2021towards}; however, in this work it refers specifically to transforming metamodels to \texttt{RDF} graphs to provide unified representation compatible with semantic technologies, a concept still largely underexplored in multi-layered DTs. Gil et al. \cite{gil2025architecture} address multi-DT coupling by applying semantic lifting to transform DT configurations and co-simulation setups into knowledge graphs, enabling querying and reasoning for consistency. Demonstrated on a three-tank system and a robotic manufacturing cell, the approach enhances reuse and monitoring. Lifting metamodel to ontology supports semantic ground and enables semantic interoperability, reasoning, and integration of evolving knowledge. Study in \cite{kappel2006lifting} introduces \emph{lifting metamodels to ontologies}, making implicit concepts explicit to support interoperability across modeling languages and facilitating integration of heterogeneous models. This ontology-based formalization enables consistent reasoning, improved conceptual matching, and more robust model integration in complex systems.

\textit{Ontology and Knowledge Graph Alignment}
Ontology alignment identifies semantically equivalent concepts and relationships, while entity alignment matches corresponding entities across knowledge graphs \cite{zhu2024survey}. Numerous studies have been proposed on ontology \cite{hertling2023olala,he2023exploring,lushnei2026large,zhang2024large,babaei2024llms4om,sousa2025complex} and knowledge graph alignment \cite{zhang2023autoalign,ramonell2023knowledge,tian2025systematic,chen2024llm}, with many of them emphasizing the growing role of \texttt{LLMs}. In contrast, heterogeneous model alignment targets semantic correspondences across different abstraction levels, i.e., model–metamodel, metamodel–ontology, and metamodel–knowledge graph. Despite their importance, these heterogeneous model alignment approaches remain largely underexplored in multi-layered DTs, particularly when ontologies and knowledge graphs are part of the conceptual framework. Study in \cite{ruckhaus2023applying} uses \emph{linked open terms} to align a public bus transport ontology with the transmodel metamodel. It achieves metamodel-ontology alignment by mapping transmodel \texttt{UML} concepts and component ontologies to ontology modules, covering agencies, routes, and journeys, ensuring semantic consistency and interoperability while addressing challenges from transmodel complexity. 

In this work, we adopt a dynamic approach to metamodel–ontology alignment within a semantic grounding pipeline, unlike prior static methods. Existing DT research primarily addresses model alignment within the same layer or formalism, while cross-level alignment (model–metamodel and metamodel–ontology) remains largely unexplored. Semantic lifting, ontologies, and knowledge graph alignment provide key mechanisms to enable DT-aware semantic grounding.

\section{DT Semantic Grounding}\label{sec3} 

DT lifecycle span design, production, operation, and end-of-life phases, requiring integration across data, models, and knowledge layers~\cite{pronost2024digital}. However, most DT implementations remain centered on data and modelling approaches, often lacking an additional semantic layer to support interoperability, consistency, and reuse~\cite{karabulut2024ontologies}. This limitation is due to heterogeneity across different abstraction layers (e.g., \texttt{JSON}, \texttt{UML}, \texttt{Ecore}, \texttt{OWL}) leading to semantic fragmentation, hindering cross-layer reasoning, interoperability and lifecycle continuity \cite{karabulut2024ontologies}. Metamodels capture system structure with \texttt{entities}, \texttt{attributes}, and \texttt{relationships}, but aligning them with ontologies is difficult due to diverse formalisms (\texttt{JSON}, \texttt{UML}, \texttt{Ecore}, \texttt{OWL}). To address this, we introduce a  semantic grounding pipeline that embeds semantics consistently across layers and lifecycle phases. The DT semantic grounding pipeline operates in four main steps: (i) \textbf{semantic alignment}, where metamodel elements are linked to domain ontology concepts to establish a shared vocabulary at design time, (ii) \textbf{semantic mapping}, where sensor and operational data is mapped to semantic concepts during deployment, sub-phase of manufacturing and (iii) \textbf{semantic evolution}, where mappings are continuously updated and maintained with incoming data and (iv) \textbf{semantic reasoning}, which enables querying, inference and consistency checking across abstraction layers during operation. At the core of this pipeline, is semantic lifting that transforms DT metamodels into \texttt{RDF} graphs and enables semantic grounding. It represents \texttt{entities} as nodes, \texttt{attributes} as \texttt{predicate-object} pairs, and \texttt{relationships} as \texttt{triples}, producing a unified, machine-interpretable representation compatible with ontologies and knowledge graphs. This process relies on metamodel structure, accompanied by some textual descriptions to provide necessary semantic context. This unified representation supports all phases of DT semantic grounding.

\begin{definition} \label{def-1}
Let \textbf{MM} (\textbf{E}, \textbf{A}, \textbf{R}) be a metamodel having entities, attributes and relations, \textbf{O} (\textbf{C}, \textbf{P}, \textbf{\textbf{$\subseteq$}}) be an ontology having concepts, properties and relation, and $\textbf{D}$ be operational data. Semantic grounding consists of two functions (alignment, mapping): $\mathcal{G}=(\phi, \psi)$. Alignment is defined as $\phi~(x) \rightarrow (y, \delta): where~x=\textbf{MM}~(\textbf{E} \cup \textbf{A} \cup \textbf{R})~and~ y= \textbf{O}~(\textbf{C} \cup \textbf{P}$) and $\delta= semantic~relation (\texttt{skos:closeMatch},...)$. Mapping is defined as $\psi: \textbf{D} \rightarrow \textbf{O} ~(\textbf{C}$), where operational data is mapped to ontology concepts. Semantic grounding enables interoperability and traceability such that domain knowledge about DT models and operational data can be derived.
\end{definition}
Here, we focus on semantic alignment and mapping by formalizing semantic grounding in Definition~\ref{def-1} and illustrating its significance through Example~\ref{exmp-1}.
\begin{example} \label{exmp-1}
Suppose that $\mathcal{D}$ includes \texttt{HumiditySensor01.value=78}. Without semantic grounding, the DT interprets \texttt{78} as a raw value without explicit meaning. With semantic grounding (alignment and mapping), the sensor is first aligned to \texttt{Brick:Humidity\_Sensor}; leveraging $\textbf{O}$, the DT can then infer that it is a \texttt{sensor} measuring \texttt{humidity}, which is an \texttt{environmental quantity}.
\end{example}

Based on semantic grounding we define our objective as: \emph{Given set of heterogeneous models $\mathbf{M}_{H} = [\mathbf{D}, \mathbf{M}, \mathbf{MM}, \mathbf{O}]$, where each represents a layer in DT (see Fig.~\ref{Fig1}) and data sources which originate from the AT. Each DT element is described at different abstraction levels: an instance $\mathbf{I}$ at the model layer $\mathbf{M}$, an entity $\mathbf{E}$ at the metamodel layer $\mathbf{MM}$, and a concept $\mathbf{C}$ at the ontology layer $\mathbf{O}$. The model layer $\mathbf{M}$ is directly connected to real-world data values from the AT. Our goal is first to lift the metamodel Layer $\mathbf{MM}$ into an \texttt{RDF graph} representation, and then to semantically align entities $\mathbf{E}$ in $\mathbf{MM}$ with concepts $\mathbf{C}$ in $\mathbf{O}$ through \enquote{equivalent-to} correspondences, ensuring semantic and structural consistency across the DT lifecycle.} Following the simple knowledge organization system (\texttt{SKOS}) framework, mappings use \texttt{skos:exactMatch}, \texttt{skos:closeMatch}, and \texttt{skos:broadMatch}; this work focuses on \texttt{skos:exactMatch} to ensure strict equivalence for interoperable DT integration.

\section{DT Usecase: Indoor Air Quality}\label{sec4}
\begin{figure}[h]
    \centering
    \begin{subfigure}[b]{0.48\textwidth}
        \includegraphics[width=3in, height=1.8in]{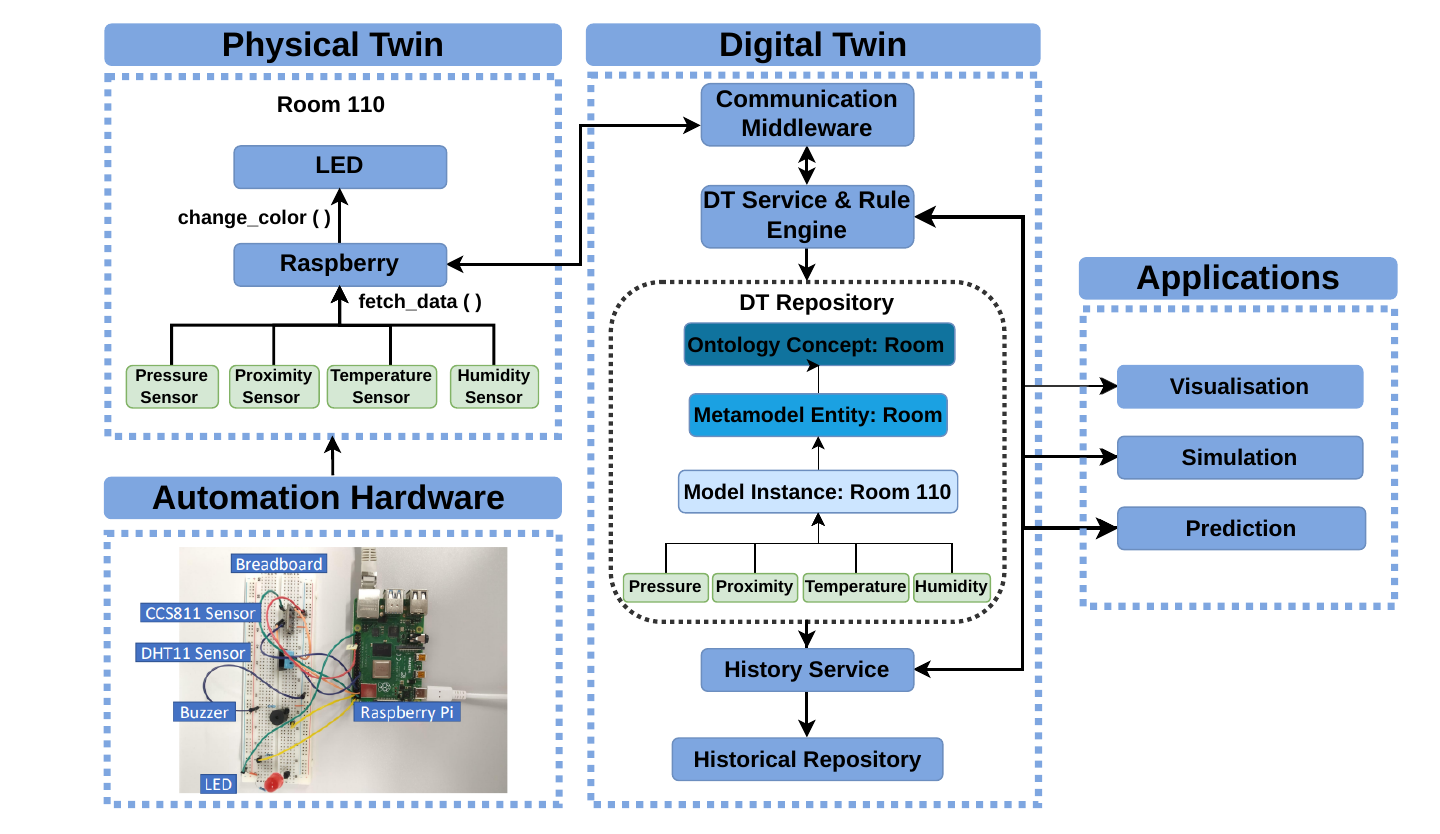}
        \caption{DT Pipeline}
        \label{Fig2a}
    \end{subfigure}
    \hfill
    \begin{subfigure}[b]{0.48\textwidth}
        \includegraphics[width=2.5in, height=1.8in]{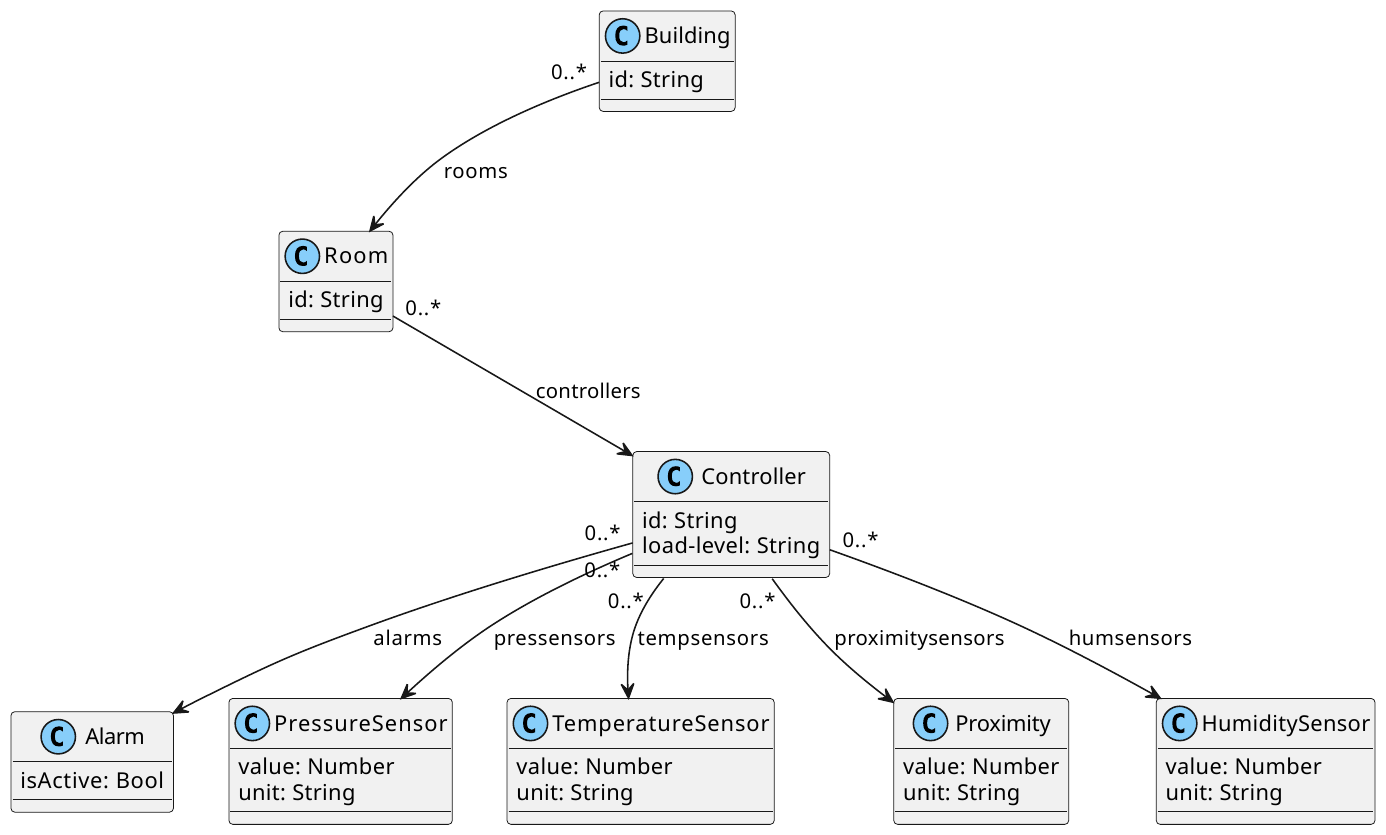}
        \caption{Exemplar Metamodel}
        \label{Fig2b}
    \end{subfigure}
    \caption{DT Usecase - Indoor Air Quality}
    \label{Fig2}
\end{figure}

This section presents an \emph{air quality management} use case, illustrating a DT for monitoring, alarms, and data-driven optimization. The metamodel (Fig.~\ref{Fig2b}) links room controllers with different sensors, as well as alarms. Based on \cite{govindasamy2021air}, DT pipeline is extended in Fig.~\ref{Fig2a}. Each room includes controller connected to sensors that stream data to DT, with alarms triggered for poor conditions.
We do not have any physical setup for physical sensor and actuators, so we integrated and mapped an \emph{air quality dataset}\footnote{\url{https://doi.org/10.17605/OSF.IO/BAEW7}} (2.5M samples from sensors)~\cite{anik2025comprehensive} to study DT-aware semantic grounding. We employ a case-based generalization strategy~\cite{wieringa2015six} to elicit the requirements in three steps: (i) \textbf{usecase positioning:} real-time monitoring with DTs in buildings, integrating sensor data (e.g., temperature, humidity) for real-time alarms, generalized to layered DT settings (data, models, metamodels, ontologies); (ii) \textbf{modeling abstractions:} domain and functional models capturing entities, relationships, and sensor-to-DT flow propagating across different abstraction layers (Fig.~\ref{Fig2a} and \ref{Fig2b}); (iii) \textbf{requirement elicitation:} structural adaptation across layers, ensuring scalable data handling, model evolution, metamodel extensibility, and semantic interoperability.

\section{Approach}\label{sec5}
This section outlines building and aligning multi-layered DTs for indoor air quality using flexible modelling framework (JSMF\cite{sottet2016jsmf}\footnote{\url{https://js-mf.github.io/}})
and semantic technologies.

\subsection{JavaScript Modelling Framework (JSMF)}\label{sec4a}
\texttt{JSMF}, inspired by \emph{Eclipse Modeling Framework (EMF)}, supports flexible modeling through partial conformance, dynamic typing, and iterative evolution \cite{sottet2016jsmf}. This suits multi-layered DTs, where data, models, and metamodels are structured in layers. We implement a multi-layered DT in \texttt{JSMF}, linking data to models conforming to the metamodel to ensure consistency. \texttt{JSMF} also enables incremental updates and structural adaptation. Further details are available here\footnote{\url{https://github.com/faimaAbbasi/dt-metamodel-semantic-grounding.git}}
 \cite{sottet2016jsmf}. For matching task, each metamodel class is annotated with a brief context-specific description using \texttt{JSMF}.

\subsection{Semantic and Structure-aware Metamodel Ontology Matching (SSM-OM)}\label{sec4b}
This subsection presents semantic alignment function ($\phi$) in multi-layered DTs by linking a source metamodel with a domain ontology for consistency and interoperability. We propose a hybrid approach, inspired by \cite{hertling2023olala}, combining graph-based embeddings \cite{muzammal2020decentralised,abbasi2019ensemble,abbasi2024snca}, lexical similarity \cite{giabelli2022embeddings}, and \texttt{LLM} reasoning \cite{amini2024towards} to align metamodel with a target ontology (e.g., \texttt{Brick \footnote{\url{https://brickschema.org/}}}
). We describe our approach in following steps: 
\begin{enumerate}
    \item \textbf{Semantic Lifting:} 
    To enable semantic grounding, we semantically lift metamodels into \texttt{RDF} graphs for a unified, standard and interoperable representation. A domain metamodel (in \texttt{JSON}), extracted from \texttt{JSMF}, is transformed into a source \texttt{RDF} graph $\mathcal{G\textsubscript{S}}$. Classes are represented as \texttt{rdfs:Class}, hierarchies as \texttt{rdfs:subClass}, and annotated with \texttt{rdfs:label} and \texttt{rdfs:description} (can be missing in classical metamodel). Attributes are modeled as \texttt{rdf:property}, either as literals or references between classes. This \texttt{RDF} graph unifies representation for ontologies, making alignment a \texttt{graph-graph} matching task that preserves semantics and structure.
    This semantic lifting yields the source triple set $\mathcal{T}_s$ as defined in Eq ~\ref{eq.1}:
    
    \begin{equation}\label{eq.1}
    \begin{split}
    \mathcal{T}_s ={} & \{ (c, \mathsf{rdf:type}, \mathsf{rdfs:Class}) \mid c \in \text{Metamodel Classes} \} \\
    & \quad \cup \{ (p, \mathsf{rdfs:domain}, c), (p, \mathsf{rdfs:range}, r) \}
    \end{split}
    \end{equation}
    
    \texttt{(p,rdfs:domain,c)} indicates that class \textbf{c} that has property 
    \textbf{p}, whereas \texttt{(p,rdfs:range,r)} defines its value type, either reference (class) or primitive (datatype). 
    This structure enables semantic alignment, and a corresponding triple set $\mathcal{T}_t$, is generated for the target ontology as depicted in Eq \ref{eq.2}.
    
    \begin{equation}\label{eq.2}
    \begin{split}
    \mathcal{T}_t ={} & \{ (c, \mathsf{rdf:type}, \mathsf{rdfs:Class}) \mid c \in \text{Ontology Classes} \} \\
    & \quad \cup \{ (p, \mathsf{rdfs:domain}, c), (p, \mathsf{rdfs:range}, r) \}
    \end{split}
    \end{equation}

    The target ontology graph $\mathcal{G\textsubscript{T}}$ is built from a domain ontology (e.g., \texttt{Brick}) in \texttt{OWL/RDFS}, capturing class hierarchies (\texttt{rdfs:subClassOf}), properties (\texttt{rdf:type}, \texttt{rdfs:domain/range}), and inter-class relationships through object properties.

    \item \textbf{Context Extraction:} Semantic context for each class $\textbf{c}$ is captured through \texttt{N-hop} neighborhood expansion using breadth first search (BFS) style traversal, producing a localized subgraph for both the source metamodel and target ontology. Eq \ref{eq.3} depicts how this context is computed:
    
    \begin{equation}\label{eq.3}
    \begin{split}
    \textbf{Context}_c^{(h)} = \bigcup_{i=1}^{h}\{ (p\textsubscript{i}, o\textsubscript{i})) \mid ( c\textsubscript{i}, p\textsubscript{i}, o\textsubscript{i}) \in \mathcal{G}\}
    \end{split}
    \end{equation}
    
    For any \texttt{RDF} graph $\mathcal{G}$, $\textbf{c} \textsubscript{i}$ is a class at hop \texttt{i}, $\textbf{p} \textsubscript{i}$ is a predicate (semantic relation, i.e., object, datatype, or ontology-level, linking classes or literals) and $\textbf{o}\textsubscript{i}$ is an object (class or datatype). Aggregating predicate-object pairs up to \textbf{h} hops, defines $\textbf{Context}_c^{(h)}$, representing the local semantic structure of class $\textbf{c}$. Each class’s local structure is described in natural language for \texttt{LLM} understanding.
    \item \textbf{Top-k Candidates:} We propose a \emph{multi-feature matching} approach combining lexical cues, structural features, \texttt{sBERT}\footnote{\url{https://www.sbert.net/docs/sentence_transformer/pretrained_models.html}} embeddings, and similarity metrics (jaro–winkler, jaccard) to identify \texttt{top-k} semantically related classes and trace evolution between a metamodel and ontology. We extract semantic and structural features from the source and target to support ontology matching.
    \begin{enumerate}
        \item \textbf{Semantic Embeddings:} \label{semantic} For each class \textbf{c} in \texttt{RDF} graph $\mathcal{G}$, a textual description aggregating its label, comments, superclass, and property-value pairs is created to generate two dense \texttt{sBERT} embeddings, i.e., a label embedding ( $\mathcal{E}\textsubscript{Label}$-lexical semantics) and a rich-context embedding ($\mathcal{E}\textsubscript{rich}$-full semantic context) using Eq \ref{eq.4}.
        \begin{equation}\label{eq.4}
        \begin{split}
        \mathcal{V} (c) = \alpha\cdot\mathcal{E}\textsubscript{Label} (\texttt{sBERT}(\texttt{label(c)})) \\ + ( 1 - \alpha ) \cdot \mathcal{E}\textsubscript{Rich}(\texttt{sBERT}(\texttt{rich(c)}) 
        \end{split}
        \end{equation}

        \texttt{rich(c)}= \texttt{label(c)} + \texttt{comment(c)} + \texttt{subClassOf(c)} + \texttt{property(c)} and it captures a class’s full semantic context. A unified embedding $\mathcal{V} (c)$ combines label and rich-context embeddings through a trade-off $\alpha\in [0,1]$. For each source metamodel and target ontology class ($\textbf{c}\textsubscript{s}$, $\textbf{c}\textsubscript{t}$), we measure the semantic proximity in vector space using Eq \ref{eq.5}.
        \begin{equation}\label{eq.5}
        \begin{split}
        \textbf{semantic-score} 
        &\rightarrow \textbf{sim}(c\textsubscript{s}, c\textsubscript{t}) \\
        &= 1 - \cos(\mathcal{V}(c\textsubscript{s}), \mathcal{V}(c\textsubscript{t}))
        \end{split}
        \end{equation}
    
        To improve alignment, a lexical boost using the jaro–winkler metric adjusts embedding similarity when class labels are highly similar, overriding the score if similarity exceeds a threshold $\theta$, depicted in Eq \ref{eq.6}.
        \begin{equation}\label{eq.6}
        \begin{split}
        \textbf{IF Jaro-winkler (L\textsubscript{s}, L\textsubscript{t})} > \theta 
        &\rightarrow 
        \textbf{semantic-score} \\
        &= \textbf{max}(\textbf{semantic-score}, 0.95)
        \end{split}
        \end{equation}
\begin{figure*}[h]
\centering
\includegraphics[width=4in, height=2in]{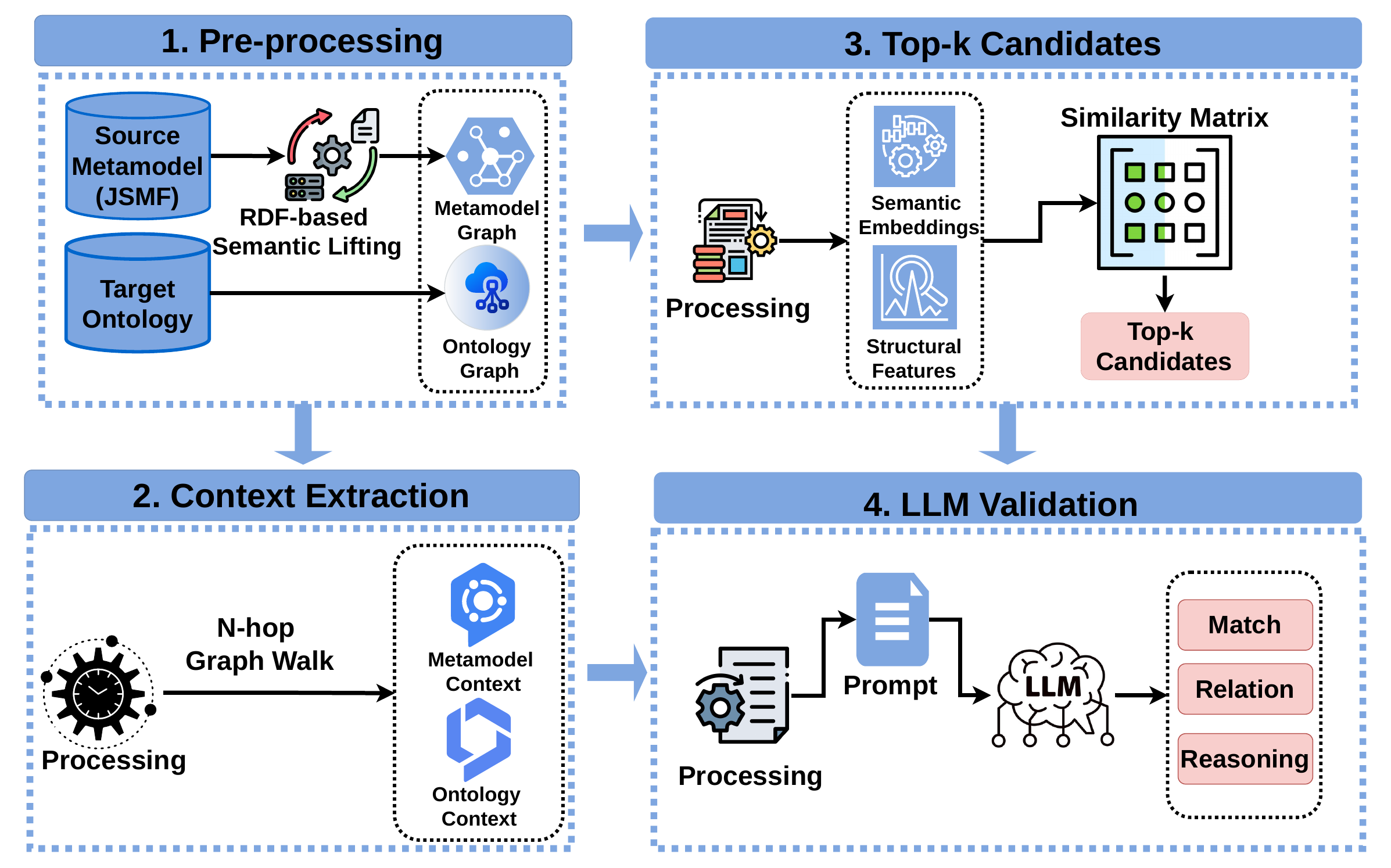}
\caption{Illustration of \texttt{SSM-OM}}
\label{Fig3}
\end{figure*}
        This rule ($\theta=0.9$) preserves near-identical lexical matches, improving early filtering, \texttt{recall}, and semantic identification.
        \item \textbf{Structural Features:}\label{structural} To capture structural features, we compute jaccard similarity over each class’s \texttt{RDF} property set \texttt{(p rdfs:domain c)} (Eq. \ref{eq.7}), providing a lightweight structural signal that complements embeddings and lexical similarity for improved alignment.
        \begin{equation}\label{eq.7}
        \begin{split}
        P(c) =\{ local\_name(p) \mid domain(p) = c \}
        \end{split}
        \end{equation}
        $\texttt{local\_name(p)}$ represents the property’s simplified label, capturing the class’s structural signature. Jaccard similarity between two classes (\textbf{c\textsubscript{s}}, \textbf{c\textsubscript{t}}), their jaccard similarity over property set is computed in Eq. \ref{eq.8}:
        \begin{equation}\label{eq.8}
        \textbf{structural\_signal} \rightarrow \textbf{Jaccard}(c_s, c_t) = 
        \frac{\left| P(c_s) \cap P(c_t) \right|}{\left| P(c_s) \cup P(c_t) \right|}
        \end{equation}
        Jaccard similarity over \texttt{RDF} property sets measures shared versus unique properties (0 if both sets are empty), capturing structural semantics to improve alignment by: (i) disambiguating classes with similar labels, (ii) revealing functional alignment across differing names, and (iii) enhancing robustness in sparse or lightweight models.
    \end{enumerate}
    We use a soft fusion strategy, combining semantic embeddings (Eq. \ref{eq.5} and \ref{eq.6}) and structural features (Eq. \ref{eq.8}) into a tunable hybrid score for each source-target class pair (Eq.~\ref{eq.9}) to enhance alignment \texttt{precision} and \texttt{recall}, which can used to track semantic and structural changes in DT lifecycle.
    \begin{equation}\label{eq.9}
    \begin{split}
    \textbf{score}(c\textsubscript{s}, c\textsubscript{t}) 
    &= \beta \cdot \textbf{semantic\_score} \\
    &\quad + (1 - \beta) \cdot \textbf{structural\_signal} \\
    \textbf{score}(c\textsubscript{s}, c\textsubscript{t}) 
    &= \beta \cdot \textbf{sim}(c\textsubscript{s}, c\textsubscript{t}) + \\ &\quad (1 - \beta) \cdot \textbf{Jaccard}(c_s, c_t)
    \end{split}
    \end{equation}
With $\beta \in [0,1]$, balancing semantic and structural signals, we compute a similarity matrix, $\textbf{S}$ $\in$ $\mathbb{R}^{n \times m}$, where $\textbf{n}$ is the number of source metamodel classes and $\textbf{m}$ is the number of target ontology classes. 
This similarity matrix contains scores for all possible class pairs computed using Eq \ref{eq.9}. Finally, we filter \texttt{top-k} candidates using Eq \ref{eq.10} based on certain threshold $\theta$.
\begin{equation}\label{eq.10}
\begin{split}
\texttt{top-k}^{\theta}(c_s) = \left\{ c_t \in \texttt{top-k}(c_s) \;\middle|\; S_{st} \geq \theta \right\}
\end{split}
\end{equation}
    \item \textbf{LLM Validation:} In our hybrid approach, the \texttt{LLM} acts as a contextual semantic evaluator, complementing graph-based methods using subclass hierarchies, property-domain links, and similarity measures (cosine embeddings, jaccard over properties) to score candidates (Eq.~\ref{eq.9}). While these methods rely on lexical and structural cues, \texttt{LLM} re-ranks \texttt{top-k} matches through local context understanding (Eq.~\ref{eq.3}), classifies relations (e.g., \texttt{skos:exactMatch}, \texttt{skos:closeMatch}, \texttt{skos:broadMatch}), and provides domain-aware justifications. A zero-shot \texttt{LLM} strategy validates \texttt{top-k} matches using contextual class representations (Eq.~\ref{eq.3}), improving alignment over purely string or embedding-based methods.
\end{enumerate}
Each metamodel class is annotated with semantic reference after successful alignment to enable semantic mapping through function $\psi$ with sensor and operational data. Fig.~\ref{Fig3} shows overview of \texttt{SSM-OM}.

\section{Evaluation}\label{sec6}
This section validates DT metamodel semantic lifting, alignment and mapping in terms of correctness, interoperability, cross-layer traceability, domain applicability and empirical performance. Matching effectiveness is assessed by comparing system-generated correspondences (\texttt{Alignment=A}) with expert references (\texttt{Reference=R}) using \texttt{precision} ($|\texttt{A} \cap \texttt{R}|/|\texttt{A}|$), \texttt{recall} ($|\texttt{A} \cap \texttt{R}|/|\texttt{R}|$), and \texttt{F1-score} ($2\times|\texttt{A} \cap \texttt{R}|/(|\texttt{A}| + |\texttt{R}|)$).

\subsection{Semantic Lifting Validation}
Semantically lifting a metamodel into \texttt{RDF} preserves its semantics and retains all structural information, enabling integration with semantic web technologies, also making it possible to reconstruct the original metamodel from \texttt{RDF}. We validate this through four tests and an alignment sensitivity analysis using \texttt{SPARQL} queries (class extraction, domain–range checks, triple counts), \texttt{JSON-RDF} reconstruction, and \texttt{SHACL} validation.
\begin{figure}[h]
    \centering
    \begin{subfigure}[b]{0.48\textwidth}
        \includegraphics[width=2.3in, height=1.7in]{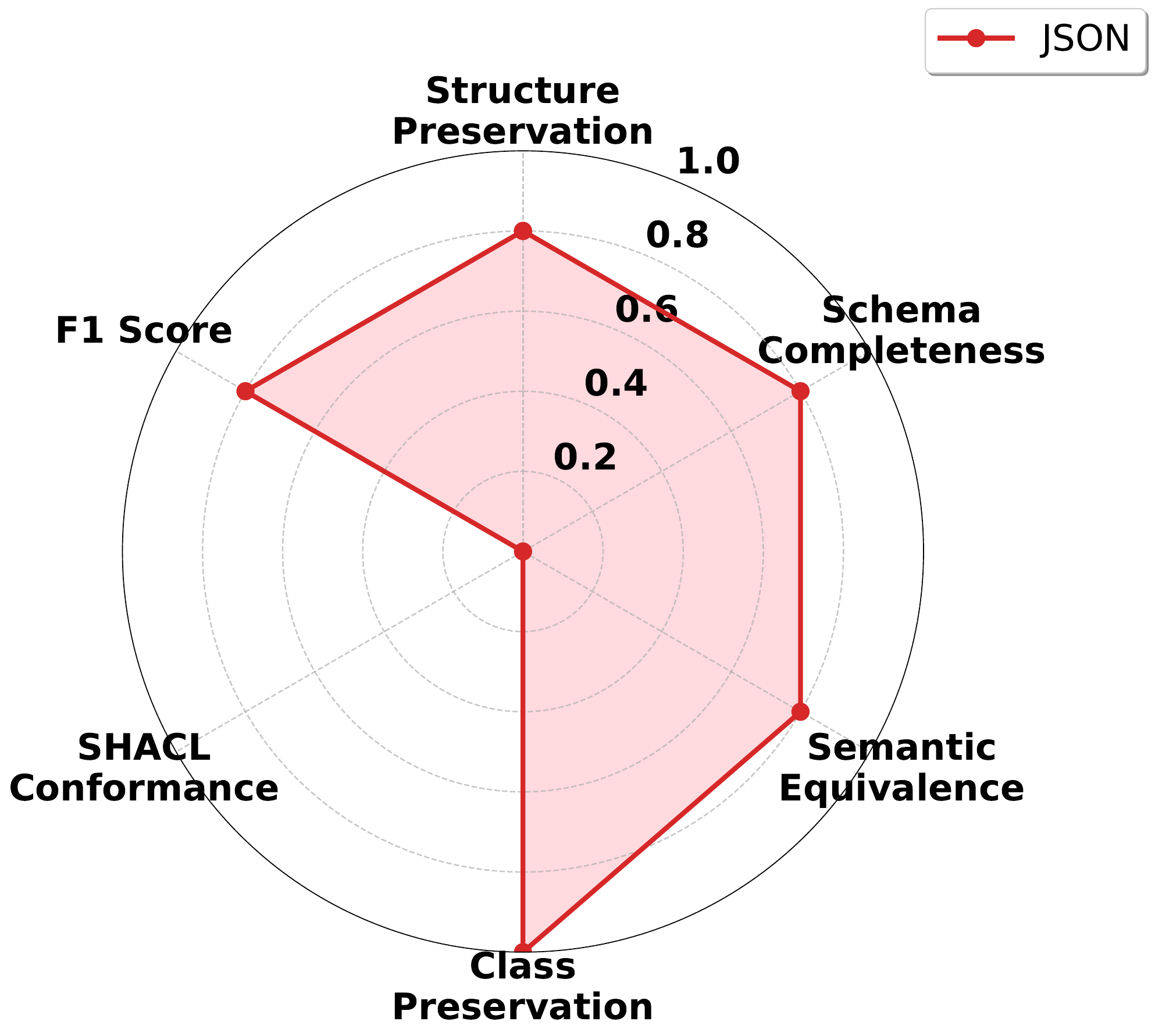}
        \caption{\texttt{JSON} Validation Metric}
        \label{Fig4a}
    \end{subfigure}
    \hfill
    \begin{subfigure}[b]{0.48\textwidth}
        \includegraphics[width=2.3in, height=1.7in]{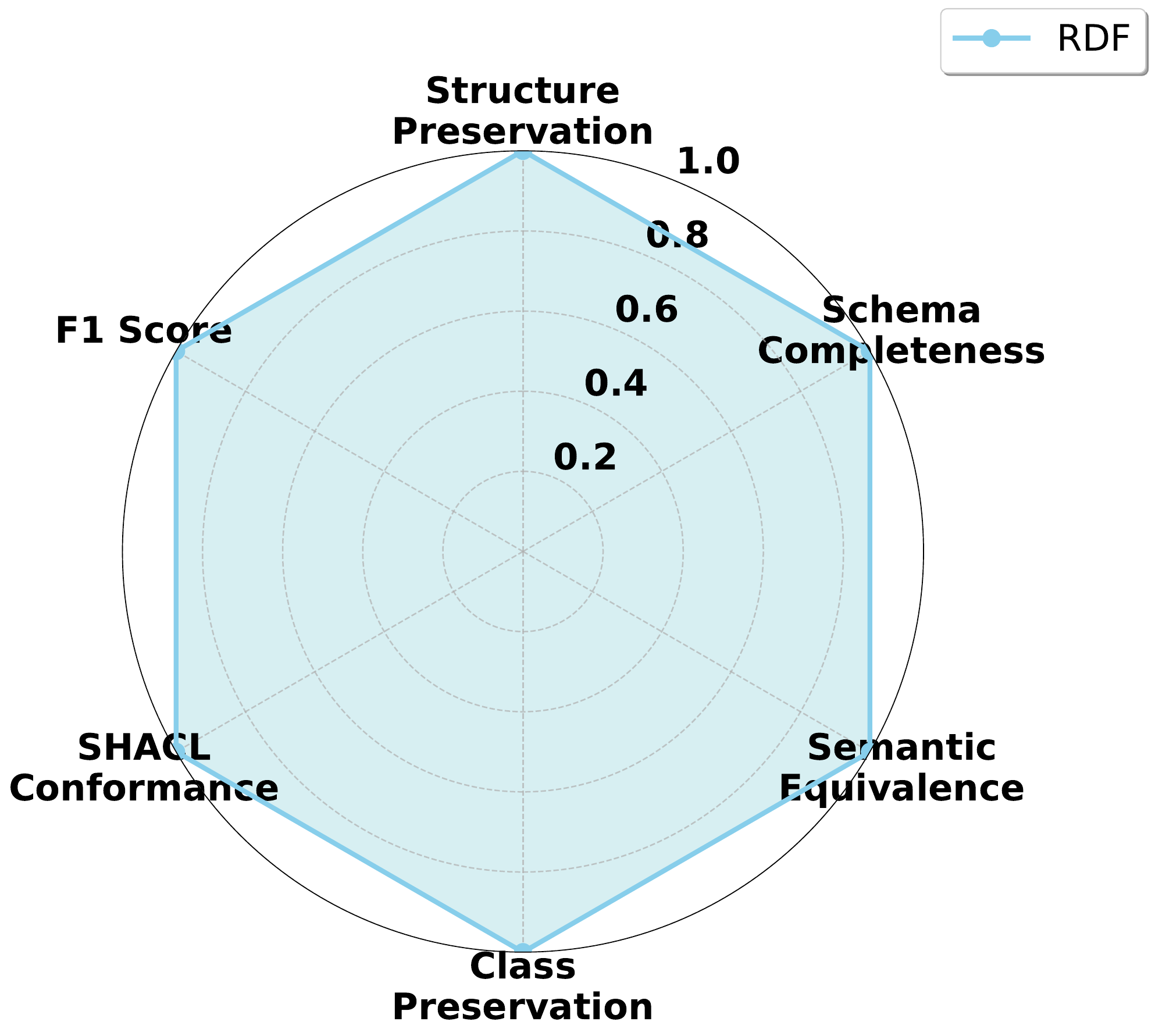}
        \caption{\texttt{RDF} Validation Metric}
        \label{Fig4b}
    \end{subfigure}
    \caption{\texttt{JSON} against \texttt{RDF} Validation Comparison}
    \label{Fig4}
\end{figure}
Specifically: (i) \textbf{round-trip validation}, reconstructing \texttt{JSON} from \texttt{RDF} using \texttt{OWL} class and property extraction with \texttt{SPARQL} checks, which preserved all classes and overall structure; (ii) \textbf{schema completeness}, ensuring full representation of the metamodel in \texttt{RDF}; (iii) \textbf{\texttt{SHACL} validation}, enforcing constraints (cardinality[\texttt{minCount=1}, \texttt{maxCount=1}]), node types, domain/range, and documentation[\texttt{rdfs:comment} checks]), which achieved zero violations; and (iv) \textbf{alignment sensitivity}, comparing \texttt{RDF} and \texttt{JSON} approaches using embeddings and lexical similarity, where \texttt{RDF} improves alignment quality by 20\% (\texttt{F1-score}: 1.0 against ~0.8). Aggregated metrics (see Fig.~\ref{Fig4}), i.e., class and structure preservation, schema completeness, semantic equivalence, and \texttt{F1-scores}, confirm that \texttt{RDF} preserves structure and semantics while outperforming \texttt{JSON}.

\subsection{DT Usecase: Indoor Air Quality}
We evaluate interoperability and cross-layer traceability in a DT use case using a unified \texttt{RDF} graph of 55{,}469 triples. To assess interoperability, we execute two complementary \texttt{SPARQL} query patterns. A metamodel-specific query (Listing~\ref{srccode:4}) returns instances but is brittle due to dependence on exact class naming. In contrast, an ontology-based query (Listing~\ref{srccode:5}) returns the same instances while leveraging \texttt{OWL} equivalence links, remaining robust to metamodel variations. This demonstrates that our approach enables external systems to query proprietary DT data using standardized vocabulary without any modification. 
\begin{lstlisting}[caption={Metamodel-specific \texttt{SPARQL} Query (Interoperability)},label={srccode:4}, mathescape=true, breaklines=true]
PREFIX mm: "<http://metamodel#>"
PREFIX dt: "<http://dt/>"
PREFIX rdf: "<http://www.w3.org/1999/02/22-rdf-syntax-ns#>"
SELECT ?instance ?value ?unit 
WHERE { ?instance rdf:type mm:TemperatureSensor .
    ?instance mm:value ?value .
    ?instance mm:unit ?unit .}
\end{lstlisting}

We evaluate cross-layer traceability in Listing \ref{srccode:6} by executing semantic propagation queries that trace connectivity across DT abstraction layers.
\begin{lstlisting}[caption={Ontology-specific \texttt{SPARQL} Query (Interoperability)},label={srccode:5}, mathescape=true, breaklines=true]
PREFIX dt: "<http://dt/>"
PREFIX mm: "<http://metamodel#>"
PREFIX brick: "<https://brickschema.org/schema/Brick#>"
PREFIX owl: "<http://www.w3.org/2002/07/owl#>"
PREFIX rdf: "<http://www.w3.org/1999/02/22-rdf-syntax-ns#>"
SELECT ?metamodelClass ?instance ?value ?unit ?brickClass 
WHERE {  BIND(mm:TemperatureSensor AS ?metamodelClass)
    ?instance rdf:type mm:TemperatureSensor .
    ?instance mm:value ?value .
    ?instance mm:unit ?unit .
    mm:TemperatureSensor owl:equivalentClass ?brickClass .
    FILTER(CONTAINS(STR(?brickClass), "Soil_Temperature_Sensor"))}
\end{lstlisting}
\renewcommand{\arraystretch}{1}
\begin{table}[H]
\centering
\caption{\textsc{Case-based Generalisation Results}}
\label{table1}
\scalebox{0.70}{%
\begin{tabular}{ll}
\hline
\textsc{Metamodel Entity}& \textsc{Ontology Concepts}    \\
\hline
http://metamodel\#Controller & https://w3id.org/rec\#Controller\\

http://metamodel\#ProximitySensor & https://w3id.org/rec\#OccupancySensorEquipment \\

http://metamodel\#Alarm & https://brickschema.org/schema/Brick\#Alarm \\

http://metamodel\#HumiditySensor & https://brickschema.org/schema/Brick\#Humidity\_Sensor \\

http://metamodel\#TemperatureSensor & 
https://brickschema.org/schema/Brick\#Soil\_Temperature\_Sensor \\

http://metamodel\#Building & https://brickschema.org/schema/Brick\#Building \\

http://metamodel\#Room & https://w3id.org/rec\#Room \\

\hline
\end{tabular}}
\end{table}
Our experiments show full semantic coverage from proprietary data to standardized ontologies, enabling DT interoperability through naming-independent querying and cross-layer traceability. Domain applicability and case-based generalization results with precise \texttt{skos:exactMatch} are summarized in Table~\ref{table1}.
\begin{lstlisting}[caption={Cross-layer Traceability},label={srccode:6}, mathescape=true, breaklines=true]
PREFIX dt: "<http://dt/>"
PREFIX mm: "<http://metamodel#>"
PREFIX brick: "<https://brickschema.org/schema/Brick#>"
PREFIX owl: "<http://www.w3.org/2002/07/owl#>"
PREFIX rdf: "<http://www.w3.org/1999/02/22-rdf-syntax-ns#>"
SELECT ?tempReading ?instance ?instanceLayer ?metamodelClass ?metamodelLayer ?ontologyClass ?ontologyLayer
WHERE {  ?instance rdf:type ?metamodelClass . 
    ?instance mm:value ?tempReading .
    ?instance mm:unit ?unit .  
    FILTER(CONTAINS(STR(?metamodelClass), "TemperatureSensor"))  
    BIND(CONCAT("Metamodel:", STR(?metamodelClass)) AS ?metamodelLayer)
    ?metamodelClass owl:equivalentClass ?ontologyClass .
    FILTER(CONTAINS(STR(?ontologyClass), "Soil_Temperature_Sensor"))
    BIND(CONCAT("Ontology:", STR(?ontologyClass)) AS ?ontologyLayer)}
\end{lstlisting}

\subsection{OAEI Testcase Performance Comparison}  
Alignment lacks ground-truth benchmarks (\texttt{R}); we evaluate strength of \texttt{SSM-OM} rather than DT application context, on four \texttt{OAEI} testcases (see Table~\ref{table2}), using \texttt{llama3} with fixed parameters: $\alpha=0.3$, $\beta=0.7$, \texttt{similarity\_threshold=0.90}, \texttt{Jaro-winkler=1}, \texttt{N\text{-}hops=5} and \texttt{top-k=3}, showing negligible \texttt{LLM}-variation and competitive performance across testcases:
(i) \textbf{anatomy:} cross-species biomedical alignment (adult mouse vs. \texttt{NCI}), \texttt{SSM-OM} ranked \texttt{5/11} in \texttt{F1-score} (2022–2023);
(ii) \textbf{conference:} 21 ontology matching tasks, \texttt{SSM-OM} ranking \texttt{5/10} with above-average \texttt{F1-score} and \texttt{recall};
(iii) \textbf{common KG:} large-scale schema alignment (\texttt{NELL}, \texttt{DBpedia}, \texttt{YAGO}, \texttt{Wikidata}), \texttt{SSM-OM} ranked \texttt{1/10} and \texttt{1/9} with top \texttt{F1-score} and \texttt{recall};
(iv) \textbf{knowledge graph:} class alignment on isolated KGs (Fandom-based), \texttt{SSM-OM} ranked \texttt{1/6} with highest \texttt{recall} and \texttt{F1-score}.

\textbf{Ablation Study:} We evaluate \texttt{SSM-OM} configurations only on \texttt{common-KG} testcase to reduce cost, with other parameters fixed. Ablation results (Table \ref{table3}) show dense context-enriched embeddings provide the strongest signal. Naive averaging of lexical (\textbf{Jaro–winkler}) and structural (\textbf{Jaccard}) similarity signals can degrade performance because the measures are not independent: they partly overlap, causing double-counting, and introduce noise when either signal is unreliable (e.g., minor lexical variation or sparse structural overlap).

\renewcommand{\arraystretch}{1}
\begin{table}[H]
\centering
\caption{\textsc{Overall results of \texttt{SSM-OM} (default) against best \texttt{OAEI} systems; current results underlined}}
\label{table2}
\scalebox{0.71}{%
\begin{tabular}{llllll}
\hline
\textsc{Track [Year]}& \textsc{Test-case}                  & \textsc{System} & \texttt{Precision}  & \texttt{Recall} & \texttt{F1-score}    \\
\hline

\multirow{11}{*}{\textbf{anatomy [2023]}}  & \multirow{11}{*}{\texttt{mouse-human}} 
        & \texttt{Matcha} &  0.95 &  0.93 &  0.94 \\
        &             & \texttt{OLaLa} &  0.92 &  0.89 &  0.91 \\
        &             & \texttt{SORBETMtch} & 0.92 & 0.89 &  0.90 \\
        &             & \texttt{LogMapBio} &  0.88 &  0.91 &  0.89 \\
        &             & \underline{\textbf{\texttt{SSM-OM}}} & \underline{\textbf{0.88}} &  \underline{\textbf{0.88}} & \underline{\textbf{0.88}} \\
        &             & \texttt{LogMap} & 0.91 & 0.84 & 0.88 \\
        &             & \texttt{AMD} & 0.93 & 0.79 & 0.86 \\
        &             & \texttt{ALIN} & 0.98 & 0.75 & 0.85 \\
        &             & \texttt{LogMapLite} & 0.96 & 0.72 & 0.82 \\
        &             & \texttt{StringEquiv} & 0.99 & 0.62 & 0.76 \\
        &             & \texttt{LSMatch} & 0.95 & 0.63 & 0.76 \\
\hline
\multirow{10}{*}{\textbf{conference [2023]}} &  \multirow{10}{*}{\texttt{dbpedia}} & \texttt{Matcha} & 0.95 & 0.93 & 0.94 \\
        &         & \texttt{SORBETMtch} & 0.92 & 0.89 & 0.90 \\
        &         & \texttt{LogMap} & 0.91 & 0.84 & 0.88 \\
        &         & \texttt{LogMapBio} & 0.88 & 0.91 & 0.89 \\
        &         & \underline{\textbf{\texttt{SSM-OM}}} & \underline{\textbf{0.87}} & \underline{\textbf{0.87}} & \underline{\textbf{0.87}} \\
        &         & \texttt{AMD} & 0.93 & 0.79 & 0.86 \\
        &         & \texttt{ALIN} & 0.98 & 0.75 & 0.85 \\
        &         & \texttt{LogMapLite} & 0.96 & 0.72 & 0.82 \\
        &         & \texttt{StringEquiv} & 0.99 & 0.62 & 0.76 \\
        &         & \texttt{LSMatch} & 0.95 & 0.63 & 0.76 \\

\hline
\multirow{20}{*}{\textbf{common KG [2022]}}  & \multirow{10}{*}{\texttt{nell-dbpedia}} 
        & \underline{\textbf{\texttt{SSM-OM}}} & \underline{\textbf{0.96}} & \underline{\textbf{0.96}} & \underline{\textbf{0.96}} \\
        &         & \texttt{OLaLa} &  1.0 & 0.92 & 0.96 \\
        &         & \texttt{KGMatcher} & 1.00 & 0.91 & 0.95 \\
        &         & \texttt{Matcha}  & 1.00 &  0.81 & 0.90 \\
        &         & \texttt{ATMatcher} & 1.00 & 0.80 & 0.89 \\
        &         & \texttt{LogMap} & 0.99 & 0.80 & 0.88 \\
        &         & \texttt{LogMapKG} &  0.98 &  0.80 &  0.88 \\
        &         & \texttt{LsMatch} &  0.96 & 0.75 & 0.84 \\
        &         & \texttt{LogMapLite} &  1.00 &  0.60 &  0.75 \\
        &         & \texttt{String Baseline} & 1.00 & 0.60 & 0.75 \\
\cmidrule{2-6}
          & \multirow{9}{*}{\texttt{yago-wikidata}} 
          & \underline{\textbf{\texttt{SSM-OM}}} & \underline{\textbf{0.93}} & \underline{\textbf{0.93}} & \underline{\textbf{0.93}} \\
          &               & \texttt{KGMatcher+} & 0.99 & 0.83 & 0.91 \\
          &               & \texttt{Matcha} & 1.00 & 0.80 & 0.89 \\
          &               & \texttt{ATMatcher} & 1.00 & 0.77 & 0.87 \\
          &               & \texttt{LogMap} & 1.00 & 0.76 & 0.86 \\
          &               & \texttt{LogMapKG} & 1.00 & 0.76 & 0.83 \\
          &               & \texttt{String Baseline} & 1.00 & 0.70 & 0.82 \\
          &               & \texttt{LogMapLite} & 1.00 & 0.70 & 0.81 \\
          &               & \texttt{LsMatch} & 0.96 & 0.63 & 0.76 \\
\hline
\multirow{12}{*}{\textbf{knowledge graph [2022]}}  & \multirow{6}{*}{\texttt{memoryalpha-stexpanded}}
        & \underline{\textbf{\texttt{SSM-OM}}} & \underline{\textbf{0.95}} & \underline{\textbf{0.95}} & \underline{\textbf{0.95}} \\
        &             & \texttt{ATMatcher} & 0.83 & 0.71 & 0.77 \\
        &             & \texttt{LogMap} & 0.88 & 0.50 & 0.64 \\
        &             & \texttt{OLaLa} & 1.00 & 0.35 & 0.53 \\
        &             & \texttt{LsMatch} & 1.00 & 0.29 & 0.44 \\
        &             & \texttt{String Baseline} & 1.00 & 0.29  & 0.44 \\
\cmidrule{2-6}
        & \multirow{6}{*}{\texttt{starwars-swtor}}  
    & \underline{\textbf{\texttt{SSM-OM}}} & \underline{\textbf{0.96}} & \underline{\textbf{0.96}} & \underline{\textbf{0.96}} \\
        &             & \texttt{ATMatcher} & 1.00 & 0.87 & 0.93 \\
        &             & \texttt{KGMatcher} & 1.00 & 0.87 & 0.93 \\
        &             & \texttt{String Baseline} & 1.00 & 0.80 & 0.89 \\
        &             & \texttt{OLaLa} & 0.92 & 0.80 & 0.86 \\
        &             & \texttt{LogMap}  & 1.00 & 0.73 & 0.85 \\
\hline
\end{tabular}}
\end{table}
\renewcommand{\arraystretch}{1.1}
\begin{table}[H]
\centering
\caption{\textsc{Impact of Different \texttt{SSM-OM} Configuration}}
\label{table3}
\scalebox{0.75}{%
\begin{tabular}{lllllll}
\hline
 &  \multicolumn{3}{c}{\texttt{nell-dbpedia}}  &  \multicolumn{3}{c}{\texttt{yago-wikidata}} \\ \hline
 \texttt{Configuration} & \texttt{Precision} & \texttt{Recall} & \texttt{F1-Score}  & \texttt{Precision} & \texttt{Recall} & \texttt{F1-Score} \\
\hline
Embeddings (\texttt{sBERT})  & 0.94  & 0.94 & 0.94 & 0.90 & 0.90  & 0.90    \\
$\oplus$  Lexical Boost (\textbf{Jaro-winkler})    & 0.82 & 0.82 & 0.82 & 0.87 & 0.87  & 0.87    \\
$\oplus$  Structure (\textbf{Jaccard})    & 0.87 & 0.87 & 0.87 & 0.89 & 0.89 & 0.89  \\
$\oplus$  \texttt{LLM} Validation & 0.96  & 0.96  & 0.96  & 0.93 & 0.93 & 0.93 \\
\hline
\end{tabular}}
\end{table}

Nevertheless, both are retained to capture complementary aspects of change in multi-layered DT, where neither view alone is sufficient. We apply \texttt{LLM} validation to selectively filter and re-weight similarity judgments.


\section{Discussion}\label{sec7} 
In this section, we qualitatively discuss semantic grounding of DT metamodels using \texttt{RDF} graphs. We validate it through: (i) \texttt{RDF}-based semantic lifting experiments confirming correctness and handling heterogeneity in multi-layered DT (Fig.~\ref{Fig4}), (ii) a DT usecase demonstrating interoperability, cross-layer traceability and generalization (Table~\ref{table1}), and (iii) comparisons with \texttt{OAEI} benchmark test cases providing general performance of \texttt{SSM-OM} (Table~\ref{table2}), consistently ranking in the top five without task-specific tuning. \texttt{LLM}-based systems scale poorly with ontology size (e.g., $>10K$ triples), limiting full graph processing. Evaluation focuses on ontology–ontology concepts due to missing ground truth \texttt{R}, which remains valid as \texttt{OAEI} ontologies derive from heterogeneous metamodels (e.g., \texttt{OWL}, \texttt{OBO}), reflecting inter-layer heterogeneity. \texttt{SSM-OM} is limited to metamodels with textual descriptions and \texttt{RDF}-representable entities; \texttt{LLMs} struggle with full \texttt{RDF} graphs due to complexity and token limits, performing better on individual classes with clearer local context.

\section{Conclusion}\label{sec8}
We present a pipeline for semantically grounding DT metamodels using \texttt{RDF} graphs. Three core concepts are used: (i) a multi-layered DT implementation using \texttt{JSMF} constructs; (ii) lifting DT metamodels into \texttt{RDF} graphs for unified representation enabling semantic alignment and mapping; and (iii) a graph-based alignment method (\texttt{SSM-OM}) to map lifted metamodels to domain ontologies. We validate the approach in terms of correctness, interoperability, cross-layer traceability, domain applicability and general performance. Future work will extend the pipeline to support semantic evolution, addressing structural and behavioral drift in multi-layered DTs.
\subsection*{Acknowledgment}
Supported by the Luxembourg National Research Fund (FNR), project MDDT-SD (C22/IS/17153694).

\bibliographystyle{splncs04}
\bibliography{ref}
\end{document}